\documentclass[sigconf,10pt,nonacm]{acmart}

\acmDOI{}
\acmISBN{}
\acmYear{2024}
\copyrightyear{2024}
\setcopyright{none}
\acmPrice{}

\usepackage{algorithm}
\usepackage{algpseudocode}

\newcommand{\algorithmicnot}{\textbf{not}}
\algdef{SE}[IF]{IfNot}{EndIf}[1]{\algorithmicif\ \algorithmicnot\ #1\ \algorithmicthen}{\algorithmicend\ \algorithmicif}

\usepackage{supertabular}
\usepackage{shellesc}

\begin{document}

\title{Head-First Memory Allocation on Best-Fit with Space-Fitting}

\author{Adam Noto Hakarsa}
\affiliation{%
  \institution{Harvard University}
  \city{Cambridge}
  \state{MA}
  \country{USA}
}
\email{adb819@g.harvard.edu}

\renewcommand{\shortauthors}{F. Last1 et al.}

\begin{abstract}
Although best-fit is known to be slow, it excels at optimizing memory space utilization. Interestingly, by keeping the free memory region at the top of the memory, the process of memory allocation and deallocation becomes approximately 34.86\% faster while also maintaining external fragmentation at minimum.
\end{abstract}


\maketitle

\sloppy

\section{Introduction}

Memory management is a fundamental part of any robust operating system \cite{Ting1976}. Architecturally, this is because CPU must cooperate with the memory \cite{Liu2011}, which is currently the fastest and cheapest storage medium next to the register. Even if such architectural limitations were absent, memory management would still be essential because it is generally impractical for any program to accurately estimate its memory needs in advance, hence the need for dynamic memory allocation. Speed, too, matters as 30-60\% of programs spend their execution time in allocating dynamic memory \cite{HASAN2005, Zorn1992, Berger2001}. The fact that main memory is limited in size further underscores the importance of memory management.

From an OS perspective, memory management typically involves a linked list to keep track of memory usage. Several dynamic memory management algorithms, simply referred to as \textit{allocator}, operate on this large chunk of space, among which the most commonly reviewed are first-fit, next-fit, best-fit, and quick-fit. 

The first-fit algorithm scans the memory from the beginning until it finds the first free segment large enough for the request \cite{Liu2011}. The next-fit algorithm operates similarly, except it starts scanning from the point where it stopped last time \cite{Liu2011}. In contrast, the best-fit algorithm searches the entire list to find the smallest free segment that meets the request \cite{Liu2011}. Alternatively, the quick-fit algorithm maintains lists of memory segments of specific sizes and, upon receiving an allocation request, searches the list of segments closest in size to the requested one \cite{Liu2011}.

The choice of an allocation algorithm is a compromise between efficient use of memory and low allocation overhead \cite{Shore1975}. This is why first-fit and best-fit are popular, especially since it does not require computing statistical distributions or maintaining an extraneous data structure which requires an additional time and space.

\section{Background}

Due to its simplicity, the first-fit and next-fit algorithms may result in memory waste through internal fragmentation, which occurs when the allocated block is larger than the requested size, leaving some space within the block unused. Consequently, best-fit or quick-fit algorithms are often preferred because they aim to allocate the smallest possible block. However, these algorithms still suffer from external fragmentation. This type of fragmentation can prevent the operating system from allocating memory even if sufficient free space exists. Techniques such as compaction, coalescing, segmentation, and paging attempt to address this issue. Despite this, best-fit is effective in optimizing the use of limited memory space. Therefore, we aim to explore a simple technique to expedite the best-fit algorithm.

\section{Algorithm}

Our allocator does not have a minimum allocation size, although blocks must always be located at addresses that are multiples of eight (double word) to ensure compatibility with systems such as Sun workstations \cite{HASAN2005}. Each allocated memory block includes a bookkeeping structure that records essential data. We have minimized the size of this bookkeeping structure to 16KB, storing only key information: whether the block is free, the block’s owner process ID, the block’s addressable space size, and a link to the previous block in the chain. This link is necessary because, although we can move forward using pointer arithmetic, we cannot move backward since we do not know the size of the block to the left.

It is important to note that the best-fit algorithm alone can lead to increased external fragmentation. To address this issue, we employ functions such as \texttt{SpaceFit} and \texttt{ChunkUp}, which we will discuss in detail later. The allocation process is managed by a function called \texttt{Create}. We have observed that a small change in the \texttt{Create} implementation can significantly speed up the memory allocation process, which we will demonstrate later.

\subsection{Allocation}

The process to assign an area in memory to a program is called (storage) allocation \cite{Ting1976}. Such a process may fail for reasons such as the lack of free block to accommodate the request.

\begin{algorithm}
\caption{Reserve a block without regard to head-first}
\label{alg:non-head-first-create}
\begin{algorithmic}[1]
\Function{Create}{reqSize}
  \State reqSize $\gets$ DOUBLEALIGN ( reqSize ) 
  \State *freeBlock $\gets$ Find ( reqSize ) 
 
  \If{no free block}
   \State freeBlock $\gets$ Stitch ( reqSize ) 
  \EndIf

  \State \Return {if still no free block}

  \If{ block's size is bigger than request }
   \State freeBlock $\gets$ ChunkUp ( this , reqSize ) 
   \State freeBlock $\gets$ SpaceFit ( this , reqSize ) 
  \EndIf

  \State \Return freeBlock
\EndFunction
\end{algorithmic}
\end{algorithm}

\begin{algorithm}
\caption{Reserve a block on head-first}
\label{alg:head-first-create}
\begin{algorithmic}[1]
\Function{Create}{reqSize}
  \State reqSize $\gets$ DOUBLEALIGN ( reqSize ) 
  \State *freeBlock $\gets$ Find ( reqSize ) 
 
  \If{no free block}
   \State freeBlock $\gets$ Stitch ( reqSize ) 
  \EndIf

  \State \Return {if still no free block}

  \If{ block's size is bigger than request }
   \State freeBlock $\gets$ SpaceFit ( this , reqSize ) 
  \EndIf

  \State \Return freeBlock
\EndFunction
\end{algorithmic}
\end{algorithm}

The two algorithms are evidently very similar to each other, except that in the head-first algorithm, we do not call \texttt{ChunkUp}, unlike in Algorithm \ref{alg:non-head-first-create} at line 10. Additionally, the \texttt{DOUBLEALIGN} function (or macro) ensures that memory blocks are aligned on a double-word boundary.

The \texttt{ChunkUp} algorithm simply partitions a block into 2 smaller block, as long as the partition results in a usable memory block that can fit the initial request.

\begin{algorithm}
\caption{Partition a block into 2 smaller blocks}
\begin{algorithmic}[1]
\Function{ChunkUp}{*block, reqByteSize}
\State \Return {block if it is not free}
  \State {calculate halfed size with bookkeeping overhead}
  \State \Return {block if halfed size too small}
  \State

  \State {divide block into two}
  \State reconfigure block links
  \State \Return block
\EndFunction
\end{algorithmic}
\end{algorithm}

With or without \texttt{ChunkUp}, we employ space-fitting to reduce external fragmentation by calling \texttt{SpaceFit}. This function calculates the extra, redundant bytes and then transfers them to any possible adjacent block or carves a new one if possible.

\begin{algorithm}
\caption{Prevent wasting memory bytes}
\begin{algorithmic}[1]
\Function{SpaceFit}{*block, reqSize}
  \State \Return { block if no extra bytes }
  \State
 
  \If{ the next block is free }
    \State enlarge the next block
    \State shrink current block
    \State reconfigure the links
    \State \Return the shrank block
  \ElsIf{ the previous block is free }
    \State enlarge the previous block
    \State shrink current block
    \State reconfigure the links
    \State \Return the shrank block
  \ElsIf{ extra bytes > (3 * overhead struct size) }
    \State create a block to contain extra bytes first
    \State recreate the shrank block
    \State reconfigure the links
    \State \Return the shrank block 
  \EndIf

  \State \Return block 
\EndFunction
\end{algorithmic}
\end{algorithm}

The space-fitting process operates as follows: after identifying a block that is significantly larger than required, any extra bytes are transferred to the right-hand block if it is free. If only the left-hand block is free, the extra bytes are transferred there. In the rare case where neither block is free, the block will divide itself as long as no resulting block has zero addressable space. If none of these options are viable, the block remains as-is.

Lastly, \texttt{Stitch} is a simple function that attempts to coalesce free blocks from the bottom to the top. This process can result in a larger block by combining several free blocks. Without coalescing, it is possible that a user might request memory that no single block can serve unless some blocks are stitched together.

\subsection{Deallocation}

The \texttt{Free} function as demonstrated by Algorithm \ref{alg:free-func} returns a status indicating whether the block is freed (\texttt{FREED}), un-freed because it wasn't allocated to begin with (\texttt{UNALLOCATED}), or un-freed because the block is owned by another process (\texttt{SEGFAULT}). It accepts \texttt{ptr} which points to a region of memory previously allocated by the \texttt{malloc} function.

\begin{algorithm}
\caption{Free a memory block given its pointed data}
\label{alg:free-func}
\begin{algorithmic}[1]
\Function{Free}{*ptr, isForced}
  \State \Return UNALLOCATED if ptr is NULL

  *this $\gets$ memory block pointed to by ptr
 
  \If{ this.bytes == ptr }
    \State \Return UNALLOCATED if this.isFree
    \State \Return SEGFAULT if not owned $\&$ not isForced
    \State indicate this block is free
    \State merge with the previous block if possible
    \State merge with the right block if possible
    \State reconfigure links
    \State \Return FREED
  \EndIf
 
  \State \Return UNALLOCATED 
\EndFunction
\end{algorithmic}
\end{algorithm}

\section{Simulation}

When the memory is initialized, its underlying linked list will be laid out in the following manner:

\begin{table}[!ht]
    \centering
    \caption{Memory state upon initialization}
    \begin{tabular}{|c|c|c|c|c|}
    \hline
        i & Address & Left Addr. & Free? & Size \\ \hline
        0 & 0x143000010 & 0x0 & yes & 8388584 \\ \hline
        8388600 & 0x143800008 & 0x143000010 & yes & 8388600 \\ \hline
    \end{tabular}
\end{table}

The position, denoted as \texttt{i} starts counting from zero. The \texttt{address} represents the memory address accessible by the user. While \texttt{i} accounts for the bookkeeping struct, the \texttt{address} does not; thus, it refers to the addressable allocated memory that can be read, written, and freed. The \texttt{left addr.} indicates the memory block to its left-hand side in the chain. The \texttt{free} field indicates whether a block is currently reserved or not. The \texttt{size} field reports the size of the addressable bytes. When aggregating the \texttt{size}, the sum will be smaller than the total free memory space in the kernel-fresh state due to overhead from the bookkeeping structs created for each memory block.

It is easy to distinguish head-first from otherwise the non head-first allocation. In the head-first implementation, the unallocated region of the memory can be seen at the top as evident from table \ref{top-head-first-table}.

\begin{table}[!ht]
    \centering
    \caption{Head-first layout}
    \begin{tabular}{|c|c|c|c|c|}
    \hline
        i & Address & Left Addr. & Free? & Size \\ \hline
        0 & 0x12e000010 & 0x0 & no & 8 \\ \hline
        24 & 0x12e000028 & 0x12e000010 & yes & 16776976 \\ \hline
        16777016 & 0x12effff48 & 0x12e000028 & no & 16 \\ \hline
        16777048 & 0x12effff68 & 0x12effff48 & yes & 128 \\ \hline
        16777192 & 0x12efffff8 & 0x12effff68 & no & 8 \\ \hline
    \end{tabular}
    \label{top-head-first-table}
\end{table}

On a non head-first implementation, the unallocated region is at the bottom of the list, as evident from table \ref{non-head-first-table}.

\begin{table}[!ht]
    \centering
    \caption{Non head-first layout}
    \begin{tabular}{|c|c|c|c|c|}
    \hline
        i & Address & Left Addr. & Free? & Size \\ \hline
        0 & 0x13d800010 & 0x0 & no & 8 \\ \hline
        24 & 0x13d800028 & 0x13d800010 & no & 8 \\ \hline
        48 & 0x13d800040 & 0x13d800028 & yes & 128 \\ \hline
        192 & 0x13d8000d0 & 0x13d800040 & no & 16 \\ \hline
        224 & 0x13d8000f0 & 0x13d8000d0 & yes & 8388360 \\ \hline
        8388600 & 0x13e000008 & 0x13d8000f0 & yes & 8388600 \\ \hline
    \end{tabular}
    \label{non-head-first-table}
\end{table}

If we want to allocate 8 bytes of memory using the best-fit strategy, we would scan the linked list to find the smallest block that can accommodate at least 8 bytes. In a non head-first approach, we would split the block located at position 48 to create the required allocation.

\begin{table}[!ht]
    \centering
    \caption{Allocating 32 bytes without head-first}
    \begin{tabular}{|c|c|c|c|c|}
    \hline
        0 & 0x12c000010 & 0x0 & no & 8 \\ \hline
        24 & 0x12c000028 & 0x12c000010 & no & 8 \\ \hline
        48 & 0x12c000040 & 0x12c000028 & no & 32 \\ \hline
        96 & 0x12c000070 & 0x12c000040 & yes & 80 \\ \hline
        192 & 0x12c0000d0 & 0x12c000070 & no & 16 \\ \hline
        224 & 0x12c0000f0 & 0x12c0000d0 & yes & 8388360 \\ \hline
        8388600 & 0x12c800008 & 0x12c0000f0 & yes & 8388600 \\ \hline
    \end{tabular}
    \label{malloc-32-without-head-first}
\end{table}

However, on a head-first implementation, we don't need to traverse the list. Since the unallocated memory is at the top, we can simply request a new block that immediately fits the request, as evident from table \ref{malloc-32-with-head-first}.

\begin{table}[!ht]
    \centering
    \caption{Allocating 32 bytes with head-first}
    \begin{tabular}{|c|c|c|c|c|}
    \hline
        i & Address & Left Addr. & Free? & Size \\ \hline
        0 & 0x12e000010 & 0x0 & no & 8 \\ \hline
        24 & 0x12e000028 & 0x12e000010 & yes & 16776928 \\ \hline
        16776968 & 0x12effff18 & 0x12e000028 & no & 32 \\ \hline
        16777016 & 0x12effff48 & 0x12effff18 & no & 16 \\ \hline
        16777048 & 0x12effff68 & 0x12effff48 & yes & 128 \\ \hline
        16777192 & 0x12efffff8 & 0x12effff68 & no & 8 \\ \hline
    \end{tabular}
    \label{malloc-32-with-head-first}
\end{table}

In both implementations, a block will be merged with its right-hand or left-hand buddy whenever possible to minimizes external fragmentation. Therefore, according to Table \ref{merged-freed-32-non-head-first}, freeing the 32-byte block results in a larger block of size 128 bytes. The size is 128 bytes instead of 112 bytes because we only need one overhead struct for each memory block. Hence, any redundant bookkeeping structs get dissolved to be a part of the addressable space.

\begin{table}[!ht]
    \centering
    \caption{After merging 32-byte block on non head-first}
    \begin{tabular}{|c|c|c|c|c|}
    \hline
        i & Address & Left Addr. & Free? & Size \\ \hline
        0 & 0x149000010 & 0x0 & no & 8 \\ \hline
        24 & 0x149000028 & 0x149000010 & no & 8 \\ \hline
        48 & 0x149000040 & 0x149000028 & yes & 128 \\ \hline
        192 & 0x1490000d0 & 0x149000040 & no & 16 \\ \hline
        224 & 0x1490000f0 & 0x1490000d0 & yes & 8388360 \\ \hline
        8388600 & 0x149800008 & 0x1490000f0 & yes & 8388600 \\ \hline
    \end{tabular}
    \label{merged-freed-32-non-head-first}
\end{table}

\begin{table}[!ht]
    \centering
    \caption{After merging 32-byte block with head-first}
    \begin{tabular}{|c|c|c|c|c|}
    \hline
        i & Address & Left Addr. & Free? & Size \\ \hline
        0 & 0x11d800010 & 0x0 & no & 8 \\ \hline
        24 & 0x11d800028 & 0x11d800010 & yes & 16776976 \\ \hline
        16777016 & 0x11e7fff48 & 0x11d800028 & no & 16 \\ \hline
        16777048 & 0x11e7fff68 & 0x11e7fff48 & yes & 128 \\ \hline
        16777192 & 0x11e7ffff8 & 0x11e7fff68 & no & 8 \\ \hline
    \end{tabular}
    \label{merged-freed-32-head-first}
\end{table}

\section{Benchmark test}

Our benchmark test suite aims to execute \texttt{n} rounds of memory allocation and deallocation requests, with each allocation not exceeding 1,024 bytes. Each request is handled by a separate thread to simulate multiprocessing scenarios. We randomize both the number of bytes to allocate, and whether to allocate or deallocate at any given time. Consequently, each trial may result in a different state of the linked list, while the total CPU time remains quite consistent across different trials. It is noteworthy that the number of allocation and deallocation requests are pretty well balanced.

We record the results of executing the non head-first best-fit algorithm with space-fitting in Table \ref{non-head-first-exp}. It illustrates the number of requests performed, the execution time, the percentage of successful memory allocations and deallocations, and the total external fragmentation in bytes. The entire memory is initialized to a size of 16 megabytes.

\begin{table}[!ht]
    \centering
    \caption{Non Head-First Best-Fit Experiment Result}
    \begin{tabular}{|c|c|c|c|c|}
    \hline
        Req. & t (sec) & Malloc & Free-ed & Ex. Frag. \\ \hline
        10000 & 0.223 & 100\% & 97.53\% & 14460.82 \\ \hline
        20000 & 0.963 & 99.99\% & 99.64\% & 12127.98 \\ \hline
        30000 & 1.985 & 99.98\% & 97.71\% & 10144.12 \\ \hline
        40000 & 4.725 & 99.99\% & 99.14\% & 6438.40 \\ \hline
        50000 & 7.455 & 99.98\% & 99.41\% & 3557.71 \\ \hline
        60000 & 9.233 & 99.99\% & 98.18\% & 2067.07 \\ \hline
        70000 & 11.437 & 99.99\% & 99.46\% & 421.55 \\ \hline
        80000 & 21.942 & 79.4\% & 79.17\% & 0.00 \\ \hline
    \end{tabular}
    \label{non-head-first-exp}
\end{table}

Table \ref{head-first-exp} illustrates the experiment on head-first best-fit with space-fitting. In addition, it shows the improvement of execution time in percentage over the experiment illustrated by Table \ref{non-head-first-exp}.

\begin{table}[!ht]
    \centering
    \caption{Head-First Best-Fit Experiment Result}
    \begin{tabular}{|c|c|c|c|c|c|}
    \hline
        Req. & t (sec) & t \textsubscript{imp} & Malloc-ed & Free-ed & Ex. Frag. \\ \hline
        10000 & 0.164 & 26.46\% & 100\% & 99.49\% & 15504.29 \\ \hline
        20000 & 0.636 & 33.96\% & 99.98\% & 99.85\% & 11426.22 \\ \hline
        30000 & 1.207 & 39.19\% & 99.97\% & 98.68\% & 9554.46 \\ \hline
        40000 & 2.106 & 55.43\% & 99.99\% & 98.39\% & 7157.90 \\ \hline
        50000 & 3.507 & 52.96\% & 99.99\% & 98.54\% & 4246.84 \\ \hline
        60000 & 5.141 & 44.32\% & 99.99\% & 99.77\% & 1780.99 \\ \hline
        70000 & 9.29 & 18.77\% & 91.58\% & 92.13\% & 0.00 \\ \hline
        80000 & 12.625 & 42.46\% & 84.17\% & 83.24\% & 0.00 \\ \hline
    \end{tabular}
    \label{head-first-exp}
\end{table}

Demonstrably, the same best-fit mechanism produces different results under different operation modes, namely head-first and non head-first. We observe a significant improvement in execution time with the head-first mechanism, while also maintaining, if not improving, algorithm effectiveness.

\section{Future works}

We compare head-first versus non head-first specifically for the best-fit algorithm. We can investigate whether similar benefits apply to other memory allocation algorithms such as first-fit, next-fit, worst-fit, as well as other algorithms like fast-fits \cite{Stephenson1983} and half-fit \cite{Ogasawara1995}. Additionally, benchmarking on real-world examples, as demonstrated in \cite{HASAN2005}, can provide further insights and practical applicability.

\section{Conclusion}

We compared two best-fit implementations that are only slightly different from one another. Our benchmark has shown that operating in head-first mode, where the free unallocated region is kept near the head of the memory, speeds up best-fit operations.

\bibliographystyle{ACM-Reference-Format}
\bibliography{systor-formatted}


\begin{thebibliography}{8}


\ifx \showCODEN    \undefined \def \showCODEN     #1{\unskip}     \fi
\ifx \showDOI      \undefined \def \showDOI       #1{#1}\fi
\ifx \showISBNx    \undefined \def \showISBNx     #1{\unskip}     \fi
\ifx \showISBNxiii \undefined \def \showISBNxiii  #1{\unskip}     \fi
\ifx \showISSN     \undefined \def \showISSN      #1{\unskip}     \fi
\ifx \showLCCN     \undefined \def \showLCCN      #1{\unskip}     \fi
\ifx \shownote     \undefined \def \shownote      #1{#1}          \fi
\ifx \showarticletitle \undefined \def \showarticletitle #1{#1}   \fi
\ifx \showURL      \undefined \def \showURL       {\relax}        \fi
\providecommand\bibfield[2]{#2}
\providecommand\bibinfo[2]{#2}
\providecommand\natexlab[1]{#1}
\providecommand\showeprint[2][]{arXiv:#2}

\bibitem[\protect\citeauthoryear{Berger, Zorn, and McKinley}{Berger
  et~al\mbox{.}}{2001}]%
        {Berger2001}
\bibfield{author}{\bibinfo{person}{Emery~D. Berger},
  \bibinfo{person}{Benjamin~G. Zorn}, {and} \bibinfo{person}{Kathryn~S.
  McKinley}.} \bibinfo{year}{2001}\natexlab{}.
\newblock \showarticletitle{Composing high-performance memory allocators}. In
  \bibinfo{booktitle}{\emph{Proceedings of the ACM SIGPLAN 2001 conference on
  Programming language design and implementation}}
  \emph{(\bibinfo{series}{PLDI01})}. \bibinfo{publisher}{ACM}.
\newblock
\urldef\tempurl%
\url{https://doi.org/10.1145/378795.378821}
\showDOI{\tempurl}


\bibitem[\protect\citeauthoryear{HASAN and CHANG}{HASAN and CHANG}{2005}]%
        {HASAN2005}
\bibfield{author}{\bibinfo{person}{Y HASAN} {and} \bibinfo{person}{M CHANG}.}
  \bibinfo{year}{2005}\natexlab{}.
\newblock \showarticletitle{A study of best-fit memory allocators}.
\newblock \bibinfo{journal}{\emph{Computer Languages, Systems \&
  Structures}} \bibinfo{volume}{31}, \bibinfo{number}{1} (\bibinfo{date}{April}
  \bibinfo{year}{2005}), \bibinfo{pages}{35–48}.
\newblock
\showISSN{1477-8424}
\urldef\tempurl%
\url{https://doi.org/10.1016/s1477-8424(04)00021-1}
\showDOI{\tempurl}


\bibitem[\protect\citeauthoryear{Liu, Yue, and Guo}{Liu et~al\mbox{.}}{2011}]%
        {Liu2011}
\bibfield{author}{\bibinfo{person}{Yukun Liu}, \bibinfo{person}{Yong Yue},
  {and} \bibinfo{person}{Liwei Guo}.} \bibinfo{year}{2011}\natexlab{}.
\newblock \bibinfo{booktitle}{\emph{UNIX Operating System}}.
\newblock \bibinfo{publisher}{Springer Berlin Heidelberg}.
\newblock
\showISBNx{9783642204326}
\urldef\tempurl%
\url{https://doi.org/10.1007/978-3-642-20432-6}
\showDOI{\tempurl}


\bibitem[\protect\citeauthoryear{Ogasawara}{Ogasawara}{1995}]%
        {Ogasawara1995}
\bibfield{author}{\bibinfo{person}{T. Ogasawara}.}
  \bibinfo{year}{1995}\natexlab{}.
\newblock \showarticletitle{An algorithm with constant execution time for
  dynamic storage allocation}. In \bibinfo{booktitle}{\emph{Proceedings Second
  International Workshop on Real-Time Computing Systems and Applications}}.
  \bibinfo{pages}{21--25}.
\newblock
\urldef\tempurl%
\url{https://doi.org/10.1109/RTCSA.1995.528746}
\showDOI{\tempurl}


\bibitem[\protect\citeauthoryear{Shore}{Shore}{1975}]%
        {Shore1975}
\bibfield{author}{\bibinfo{person}{John~E. Shore}.}
  \bibinfo{year}{1975}\natexlab{}.
\newblock \showarticletitle{On the external storage fragmentation produced by
  first-fit and best-fit allocation strategies}.
\newblock \bibinfo{journal}{\emph{Commun. ACM}} \bibinfo{volume}{18},
  \bibinfo{number}{8} (\bibinfo{date}{Aug.} \bibinfo{year}{1975}),
  \bibinfo{pages}{433–440}.
\newblock
\showISSN{1557-7317}
\urldef\tempurl%
\url{https://doi.org/10.1145/360933.360949}
\showDOI{\tempurl}


\bibitem[\protect\citeauthoryear{Stephenson}{Stephenson}{1983}]%
        {Stephenson1983}
\bibfield{author}{\bibinfo{person}{C.~J. Stephenson}.}
  \bibinfo{year}{1983}\natexlab{}.
\newblock \showarticletitle{New methods for dynamic storage allocation (Fast
  Fits)}.
\newblock \bibinfo{journal}{\emph{SIGOPS Oper. Syst. Rev.}}
  \bibinfo{volume}{17}, \bibinfo{number}{5} (\bibinfo{date}{oct}
  \bibinfo{year}{1983}), \bibinfo{pages}{30–32}.
\newblock
\showISSN{0163-5980}
\urldef\tempurl%
\url{https://doi.org/10.1145/773379.806613}
\showDOI{\tempurl}


\bibitem[\protect\citeauthoryear{Ting}{Ting}{1976}]%
        {Ting1976}
\bibfield{author}{\bibinfo{person}{Dennis~W. Ting}.}
  \bibinfo{year}{1976}\natexlab{}.
\newblock \showarticletitle{Allocation and compaction - a mathematical model
  for memory management}. In \bibinfo{booktitle}{\emph{Proceedings of the 1976
  ACM SIGMETRICS conference on Computer performance modeling measurement and
  evaluation - SIGMETRICS ’76}} \emph{(\bibinfo{series}{SIGMETRICS ’76})}.
  \bibinfo{publisher}{ACM Press}.
\newblock
\urldef\tempurl%
\url{https://doi.org/10.1145/800200.806206}
\showDOI{\tempurl}


\bibitem[\protect\citeauthoryear{Zorn and Grunwald}{Zorn and Grunwald}{1992}]%
        {Zorn1992}
\bibfield{author}{\bibinfo{person}{Benjamin Zorn} {and} \bibinfo{person}{Dirk
  Grunwald}.} \bibinfo{year}{1992}\natexlab{}.
\newblock \showarticletitle{Empirical measurements of six allocation-intensive
  C programs}.
\newblock \bibinfo{journal}{\emph{ACM SIGPLAN Notices}} \bibinfo{volume}{27},
  \bibinfo{number}{12} (\bibinfo{date}{Dec.} \bibinfo{year}{1992}),
  \bibinfo{pages}{71–80}.
\newblock
\showISSN{1558-1160}
\urldef\tempurl%
\url{https://doi.org/10.1145/142181.142200}
\showDOI{\tempurl}


\end{thebibliography}
\end{document}